\documentclass[12pt]{article}
\usepackage[letterpaper,hmargin=1in,vmargin=1in]{geometry}
\usepackage{amssymb}
\usepackage{amsmath}
\usepackage{amsfonts}
\usepackage{graphicx}
\usepackage{overcite}
\usepackage{authblk}
\renewcommand{\baselinestretch}{1}

\begin{document}
\renewcommand{\baselinestretch}{1}
\title{Multi-Scale Modeling of Coarse-Grained Macromolecular Liquids}
\author{J. McCarty}
\author{I.Y. Lyubimov}
\author{M.G. Guenza
\footnote{To whom correspondence should be addressed. Phone: (541) 346-2877. Fax: (541) 346-4643. E-mail: mguenza@uoregon.edu}}
\affil{Department of Chemistry and Institute of Theoretical Science, University of Oregon, Eugene, O.R. 97403, USA}
\date{\today}
\maketitle
\begin{abstract}
A first-principle multiscale modeling approach is presented, which is derived from the solution of the Ornstein-Zernike equation for the coarse-grained representation of polymer liquids.
The approach is analytical, and for this reason is transferable. It is here applied to determine the structure of several polymeric systems, which have different parameter values, such as molecular length, monomeric structure, local flexibility, and thermodynamic conditions. When the pair distribution function obtained from this procedure is compared with the results from a full atomistic simulation, it shows quantitative agreement. Moreover, the multiscale procedure accurately captures both large and local scale properties while remaining computationally advantageous.

\end{abstract}

\section{Introduction}
\label{SX:INTR}
\noindent 
Understanding the structure of complex fluids hinges on our ability of achieving a detailed picture of the system on an extended range of lengthscales. Many important phenomena, which are tightly dependent on the local chemical structure, occur on a large scale. Thermodynamic properties, which can be considered as characterized by a macroscopic (infinite) lengthscale change as the local, monomeric structure is modified. For example compressibility or phase diagrams of a macromolecular liquid depend on the type of polymer involved. As the physics appearing on different lengthscales is related, an effort has to be made to formally correlate information on different lengthscales.\cite{complex,jpcmreview}
 
 Macromolecular liquids are complex, given the plethora of characteristic lengthscales that emerges from their investigation. 
Two fundamental, molecular lengthscales are relevant in the study of any macromolecular liquid, namely the monomer lengthscale, $l$, which is local, and the molecular radius-of-gyration, $R_g=l \sqrt{N/6}$, where $N$ is the molecular degree of polymerization, which roughly represents the overall size of the polymer.  Because $R_g^2$ increases linearly with the number of monomers in the chain, those two lengthscales can be separated by many orders of magnitude in real macromolecular systems of practical application.

The local scale is an effective length in which complicated local information is embedded. For example, in the simplest and most conventional description of the polymer local structure, inter- and intra-molecular forces balance each other leading to an unperturbed polymer chain statistics at any lengthscale\cite{GENNES}. However, simulations show that on the local scale entropic effects and chain connectivity produce a segmental correlation hole that manifests itself as local, intramolecular  clustered, structures of monomers.\cite{baschnagel} Because the liquid is almost incompressible and monomers from the same chain tend to cluster together due to connectivity, the ideality hypothesis does not hold on the local scale.

A second effect important on the local scale is the presence of rotational energy barriers.\cite{Flory, Ferrarini} To describe liquids of semiflexible polymers, where the local dynamics is affected by conformational transitions, the bond length is modified into an effective segment length that includes the effect of hindered rotations. This lengthscale is commonly referred to as the polymer persistence length.\cite{Flory}
 
Computer simulations are useful in providing detailed information about the macromolecular systems under study. Correlation functions can be readily calculated from simulation trajectories and compared with experiments. However, simulations are limited in the range of lengthscales they can cover, as the statistics is good up to the lengthscale defined by half the size of the simulation box. Increasing the box size, increases the number of interacting particles and the number of pair potentials that need to be calculated at any timestep, dramatically slowing down the simulation. 

For a system that is ergodic, where statistical averages over an ensemble of particles is equivalent to the average of one system over many computational time steps, one could think of limiting the number of particles simulated, while still being able to recover enough statistics by extending the length of the simulation run.
Unfortunately, the problem of error build-up at each step during the simulation run makes calculated trajectories rapidly unreliable.
This is due to the fact that trajectories are derived by numerical solutions of differential equations. These solutions are chaotic in nature, and small differences due to numerical uncertainty in the initial conditions produce large fluctuations in the output trajectory, which increase with time. The extent to which simulation trajectories are sensitive to small initial errors depends on the value of the Lyapunov exponent. Shadowing trajectories, which well represent the real ones, only last for a limited number of computational steps.\cite{Frenkel, shadowing} 

Other methods have been studied to improve the maximum number of computational steps that can be reliably simulated, including hybrid methods, e.g. implicit solvent methods, that conserve the atomistic description locally while assuming a meanfield description outside a local volume.  However, these methods remain under discussion because of the errors that they can include in the procedure.\cite{Frenkel}   

In a nutshell, if the system is described at high resolution, a large number of instantaneous forces have to be calculated at each computational time step, which slows down the computational time. Computational time steps have to be short to avoid large initial errors, but because trajectories become increasingly unreliable with the number of steps, this limits the longest timescale that can be reached in the simulation, and so the quality of the simulation output\cite{Frenkel,Tisley}.

A general strategy useful for increasing the timescale and the number of particles that can be simulated is to adopt a coarse-grained description of the system. The simulation of a coarse-grained system can include a significantly larger number of particles and can extend to a much longer time scale than simulations of the same system described with atomistic resolution. This is because internal degrees of freedom and local energy barriers are averaged out,  allowing a larger elementary time step than for atomistic simulations. Moreover, in the coarse-grained description each molecule contains a number of atoms that can be considerably reduced with respect to the full atomistic descriptions, so that the number of pair interactions that have to be calculated, at each computational step between each pair of molecules, is reduced, allowing for the simulation of larger systems.

The tradeoff of coarse-grained methods is that the local information is completely lost, and also that because the local energy barriers are averaged out and internal degrees of freedom are eliminated, the simulated dynamics is orders of magnitude faster than the real one, and there are not easy methods to recover the correct dynamics\cite{Ivan}. 

Modeling of coarse-grained systems have recently received considerable attention\cite{glasses} as simulations aspire to describe more and more complicated structures relevant for engineering applications as well as biophysical systems\cite{DePablo,Klein,Voth}. The key question in developing coarse-grained molecular models is the definition of the effective potential acting between units. The general strategy pursued to solve this problem is a bottom-up approach. First,  atomistic simulations of the molecular liquid are performed. Once the coarse-grained structure is defined by lumping together adjacent atoms in the molecule, an effective numerical coarse-grained potential is obtained, which is parametrized by optimizing the agreement against properties measured in the atomistic simulation. For example, the parameters in the effective potential can be defined by the optimization of the mean-force potential, matching the forces, or by optimizing the free parameters against known physical properties.
The potential, so derived, is used as an input to a new simulation of the coarse-grained system. The properties derived from this simulation are checked for consistency against physical properties, e.g. pair distribution functions, obtained from the atomistic simulation, and the process is repeated until a fully self-consistent description is achieved. The test quantity that is optimized is in most cases the pair distribution function, or analogously the total correlation function, as from this quantity all the properties of a liquid can be directly calculated\cite{McQuarrie}.

This numerical procedure has the limitation that the resulting optimized effective potentials are specific to the chemical structure considered and of the thermodynamics conditions in which the initial atomistic simulations were performed. In principle those potentials cannot be applied to describe the system in a different thermodynamic condition, which means that the potential is not transferable. 

Several coarse-graining methods have been presented in the literature. An extremely successful approach has been the United Atom (UA) description where each moiety of type $C$, $CH$, $CH_2$, or $CH_3$ is represented as an effective unit.\cite{KremerGrest} Effective potentials between united atoms have been derived and parametrized to reproduce the properties of the system described at the atomic level. The UA description is useful because polyolefins, which constitute a great portion of the macromolecular systems of interest,   can be fully represented as a collection of  these sites.
The UA description has been extensively tested and has been proven to be extremely reliable.\cite{Paul,  Binder1, Binder2} 
 
Combined with new techniques that speed up the equilibration step, the UA description allows for MD simulations of polymer melts with long entangled chains.\cite{Theodorou}
 
While UA simulations have proved successful in reproducing the structural and dynamical properties of polymers, there is an increasing need to extend simulations to even larger scales to capture other phenomena, such as cooperative self-assembly of supermolecular  structures, dynamics of systems approaching their second order phase transitions, simulations of bulk properties, to name a few.    
Recently, we have developed an original coarse-grained method, which is analytical and represents polymers on the lengthscale of $R_g$ as interpenetrating soft colloidal particles.\cite{YAPRL,YABLEND,SAM2006} A more refined  coarse-grained description has also been derived, where each macromolecule is represented as an elastic dumb-bell of interpenerating spheres.\cite{PRE2007}
The effective potential in these descriptions is derived by solving the Ornstein-Zernike equation in the monomer and effective unit sites, where the center-of-mass (com) of the effective unit is assumed to be an auxiliary site, i.e. there is no direct correlation between monomer and com. Relaxing this approximation does not change the outcome of the coarse-graining procedure.\cite{SAM2006} Because the monomer intramolecular distribution follows a Gaussian statistics, the Ornstein-Zernike equation can be solved analytically to produce the pair distribution function for the effective units. From the latter, the effective coarse-grained potential is calculated by applying the hypernetted-chain closure approximation.\cite{McQuarrie} The advantage of our coarse-grained description is that the obtained potential is directly transferable to systems in different thermodynamic conditions and to polymers with different degree of polymerization or different chemical structure\cite{YAPRL, YABLEND, SAM2006}. Comparison against UA simulations  of several polymeric liquids and mixtures showed quantitative agreement between the total correlation functions in the two representations.

Because the trade-off of coarse-graining procedures is that the information on the local length scale is lost, techniques have been developed to   merge the coarse-grained description with local information. In the so-called ``multiscale modeling" procedures  a hierarchy of simulations are performed on the system coarse-grained at different characteristic lengthscales, and then the information obtained from those simulations is combined into one overall description, which includes all the lengthscales of interest. 

The method that we are presenting here is a multiscale modeling procedure for the simplest macromolecular liquids, i.e. of homopolymers. Here, the lengthscales of interest are the monomer and the radius-of-gyration, so that two simulations have to be combined, one at the monomer level of description, and one where polymers are represented at the $R_g$ level, or mesoscopic simulation (MS-MD). 
We perform molecular dynamic simulations of the liquid of polymers described as soft colloidal particles in a MS-MD simulation and we combined those results with the outcome of a united atom molecular dynamic (UA-MD) simulation for the local description, through our multiscale modeling procedure.
We test our procedure for a number of systems, which include polyethylene melts with different degrees of polymerization, as well as liquids of macromolecules with different monomeric structures. The procedure is optimized and shows quantitative agreement between the total correlation functions calculated by us, and the ones obtained in a test UA-MD simulation. The advantage of our procedure, with respect to running a full UA-MD simulation of the same system, is due to the gain in computational time. 

Another advantage of  our multiscale modeling procedure, with respect to other multiscale approaches, is that it is formally compatible with the first principle formalism used to coarse-grain the polymeric structure as both rely on the Ornstein-Zernike equation which includes auxiliary sites.  Because it is analytical, the coarse-graining procedure can be used not only to remove internal degrees of freedom, but also in the opposite way to reintroduce ``a posteriori" those internal degrees of freedom after the MS-MD is completed. It is shown that this approach reproduces pair correlation functions at a high computational efficiency, providing a method of extending simulations to large length scales of interest.

\section{Coarse-graining of a polymer at the radius-of-gyration lengthscale}
Our coarse-grain model, briefly revisited here, maps each macromolecule into an interacting soft-colloidal particle. We developed this model in a series of papers, where we reported coarse graining of polymer melts with different architectures as well as coarse graining of mixtures of those polymers.\cite{jpcmreview,YAPRL,YABLEND,SAM2006, JOPCM} 

The first modeling of polymers as soft colloidal particles dates back to Flory and Krigbaum,\cite{Flory1}
where repulsive monomer-monomer interactions are integrated over the
Gaussian chain distribution. The resulting effective potential at contact is repulsive and becomes stronger 
with increasing chain length or polymer density, in disagreement with simulations, scaling theories and 
renormalization group 
calculations.
Correct scaling behavior was obtained by  Hansen and coworkers who derived pair distribution functions for polymers in dilute or semi-dilute solution, starting from first principles liquid state theory.\cite{Hansen} A more phenomenological form of this approach was implemented early on to describe polymer melts and mixtures by Dautenhahn and Hall,\cite{Hall} and by
Murat and Kremer.\cite{Kremer}

The choice of the monomer and radius-of-gyration as the two length scales at which the polymer liquid is coarse-grained in our multiscale procedure is motivated by the fact that the monomer pair distribution function, and the structure of the polymer liquid, are fully determined when these two length scales are known, as shown below.

Our coarse-graining procedure starts from Curro and Schweizer's  polymer reference interaction site model (PRISM) Ornstein-Zernike equation, which for a homopolymer melt relates the monomer total pair correlation function, $h(r)$, to the direct correlation function, $c(r)$, in reciprocal space \cite{PRISM,PRISM1}
\begin{equation}
\hat{h}(k)=\hat{\omega}(k)\hat{c}(k)[\hat{\omega}(k)+\rho \hat{h}(k)] \ ,
\label{EQ:1}
\end{equation}
\noindent where $\hat{\omega}(k)$ is the intramolecular structure factor, and $\rho$ is the monomer site number density. The procedure is then extended to include the center-of-mass of the coarse-grained structures as auxiliary sites, using the formalism of Krakoviack \emph{et al.}\cite{K2002}, which starts from a matrix form of Eq.(\ref{EQ:1}). The density of auxiliary sites is given by the molecular number density, $\rho_{ch}=\rho/N$. The solution of  Eq.(\ref{EQ:1}) is obtained under the assumptions  that the direct correlation functions, $\hat{c}(k)$, only depend on interactions between real sites, while no interaction is assumed between auxiliary sites. Relaxing this condition does not modify the zeroth order result,\cite{SAM2006} which yields the equation
\begin{equation}
h^{cc}(k)=\left[\frac{\omega^{cm}(k)}{\omega^{mm}(k)}\right]^2h^{mm}(k) \ ,
\label{EQ:2}
\end{equation}

\noindent where $h^{mm}(k)$ is the total pair monomer-monomer correlation function, $\omega^{cm}(k)$ is the intrachain correlation function of monomers about the com, $\omega^{mm}(k)$ is the intramolecular monomer-monomer correlation function, and $h^{cc}(k)$ is the mesoscale com-com total correlation function. Eq.(\ref{EQ:2}) is the basis for our coarse graining approach since it links the com and the monomer site correlation functions. The com representation corresponds to a mesoscale, coarse-grained description of the polymeric liquid, where each chain is a soft colloidal particle centered on the position of the polymer com. On the other hand, the monomer site description is related to the united atom representation of the polymer chain, where each site represents an united atom.

In order to obtain an expression for the coarse-grained description, we adopt for the monomer correlation function, $h^{mm}(r)$, the PRISM thread description, in which the polymer chain is treated as an infinite thread of vanishing thickness,
\begin{eqnarray}
h^{mm}(r) & = & \frac{\xi_{\rho}'}{r} \left[\exp \left(-\frac{r}{\xi_{\rho}} \right) - \exp \left(-\frac{r}{\xi_{c}} \right)   \right] \ ,
\label{eqhmon}
\end{eqnarray}
with $\xi'_\rho=3/(\pi \rho l^3)$ with $\rho$ the monomer site density. Moreover, $\xi_\rho$ is the length scale of density fluctuations defined as $\xi^{-1}_\rho=\xi^{-1}_c +\xi'^{-1}_\rho$, and $\xi_c=R_g/\sqrt{2}$ is the length scale of the correlation hole \cite{GENNES}.
Eq.(\ref{eqhmon}) shows how the structure of the macromolecular liquid is fully specified once the monomer and the radius-of-gyration length scales are known.

By assuming Gaussian intramolecular distributions for the monomer-monomer and monomer-com correlation functions, which is a well tested assumption, Yatsenko, \emph{et al.} have shown that $h^{cc}(r)$ can be expressed in an analytical form  in terms of molecular and thermodynamic parameters as\cite{YAPRL} 
\begin{eqnarray}
h^{cc}(r)\;=\;\frac{3}{4} \sqrt{\frac{3}{\pi}} \frac{\xi'_\rho}{R_g} \left( 1-\frac{ \xi^2_c}{ \xi^2_\rho} \right)
e^{-3r^2/(4R^2_g)} -\frac{1}{2} \frac{ \xi'_\rho}{r} \left( 1-\frac{ \xi^2_c}{\xi^2_\rho} \right) ^2 e^{R^2_g/(3\xi^2_\rho)}\nonumber\\ \times \left[ e^{r/\xi_\rho} \mbox{erfc} \left( \frac{R_g}{ \sqrt{3} \xi_\rho}+ \frac{ \sqrt{3}r}{2R_g} \right)-e^{-r/\xi_\rho} \mbox{erfc} \left( \frac{R_g}{ \sqrt{3} \xi_\rho}- \frac{\sqrt{3}r}{2R_g} \right)\right]
\label{EQ:5.1}
\end{eqnarray}
where erfc$(x)$ is the complementary error function and $\xi'_\rho$ can be expressed in terms of the radius of gyration as $R_g/(2\pi\rho^*_{ch})$ with $\rho^*_{ch}\equiv\rho_{ch}R^3_g$ being the reduced molecular number density.
Although the PRISM-thread model employed at the monomer level description does not predict solvation shells, which are present in a polymeric liquid on the local scale, its description of the structure of the liquid at the center-of-mass length scale is correct. 
For example, Eq.(\ref{EQ:5.1}) recovers the isothermal compressibility of the melt at the $k=0$ limit \cite{YAPRL} and it is thermodynamically consistent. This equation effectively maps the polymer melt into a liquid of soft colloidal particles of radius $R_g$ and recovers, in its extension to the case of polymer mixtures, the formalism of Bhatia and Thornton for mixtures of colloidal particles.\cite{YABLEND,BATIA,SMPLQ}

When compared against simulation data, the analytical expression of Eq.(\ref{EQ:5.1}) reproduces well the com total correlation function obtained from UA-MD simulations for several polymeric systems that have different degrees of polymerization, different monomeric structure, and different density and temperature.\cite{SAM2006} Theoretical calculations of $h^{cc}(k)$ do not contain any adjustable parameter because the parameters entering 
Eq.(\ref{EQ:5.1}) are the ones used in the united atom simulation against which the formalism is tested.

Starting from Eq. (\ref{EQ:5.1}) and enforcing for the closure equation the hypernetted chain approximation, which is the most reliable closure for soft potentials,  the effective intermolecular potential between the coms of a pair of soft colloidal particles, $v^{cc}(r)$, is derived as\cite{YAPRL,YABLEND,SAM2006}
\begin{equation}
\beta v^{cc}(r)=h^{cc}(r)-\ln[1+h^{cc}(r)]-c^{cc}(r) \ ,
\label{EQ:6}
\end{equation}
with $\beta=(k_B T)^{-1}$ the inverse temperature of the system in Boltzman units.
The direct pair correlation function, $c^{cc}(r)$ is given in reciprocal space  in terms of $h^{cc}(k)$ as
\begin{equation}
c^{cc}(k)=\frac{h^{cc}(k)}{1+\rho_{ch} h^{cc}(k)}
\label{EQ:6.1}
\end{equation}
Substitution of Eq.\ref{EQ:5.1} and Eq.\ref{EQ:6.1} into Eq.\ref{EQ:6} gives the effective pair potential acting between two soft colloidal particles. This is the potential needed as an input to the mesoscopic simulation of the coarse-grained units and is the potential we use in our MS-MD. 

Because coarse-grained potentials result from the mapping of many-body interactions into pair interactions, through the averaging over microscopic degrees of freedom, they are parameter dependent. During the coarse-graining procedure, the potential acting between microscopic units, which is given by the Hamiltonian of the system, reduces to an effective potential, which is a free energy in the reference system of the microscopic coordinates. The coarse-grained potential so obtained contains contributions of entropic origin due to the microscopic, averaged-out degrees of freedom and is therefore state-dependent.  This can be observed in the form of the total correlation function between coarse-grained sites, Eq.(\ref{EQ:5.1}), which explicitly includes the structural parameters of the polymer, i.e. the radius-of-gyration and density screening length, as well as the thermodynamic parameter of the system number density. The temperature enters directly through  Eq.(\ref{EQ:6}) and indirectly through the molecular parameters, such as $R_g$. 

The common procedure adopted to derive the effective potential between coarse-grained units is fully numerical. Once the level of coarse-graining is defined by the selection of the effective units, the pair distribution function between effective units, $g(r)$, is numerically calculated. From $g(r)$ the potential of mean force is derived as $- k_BT \ln g(r)$, and then used as an input to a coarse-grained simulation.
By comparison against the UA simulation, corrections to the potential are obtained, and the new potential of mean-force is used as an input to a new simulation until full consistency is obtained between coarse-grained and UA simulations.
Because numerically solved coarse-grained potentials have parameters that are defined through optimization against a specific UA simulation, they apply only for systems at thermodynamic conditions close to the one of the UA simulation, so they are not transferable. Moreover, because their  form depends on the procedure used to derive them, if they are optimized against a given set of physical properties, it is not assured that they will reproduce other properties of the system. The numerical optimization of the potential against different physical properties could lead to different effective potentials, numerically optimized, even for the same system. 

In this respect, an analytical potential such as the one we derive here has the advantage of being explicitly parameter dependent, and because it applies to systems defined by different values of the thermodynamic parameters, it is fully transferable. Moreover, because it is derived from a first principle approach, its properties are well defined. However, analytical potentials, as the ones we derived here, can be obtained only for a limited number of systems, thus implying that in most cases the numerical optimization procedure is the only option in deriving the effective coarse-grained potential. In conclusion, both analytical and numerically evaluated potentials are useful as they answer complementary questions in the process of building optimized coarse-grained descriptions of macromolecular systems.

\begin{figure}[]
\centering
\includegraphics[scale=.7]{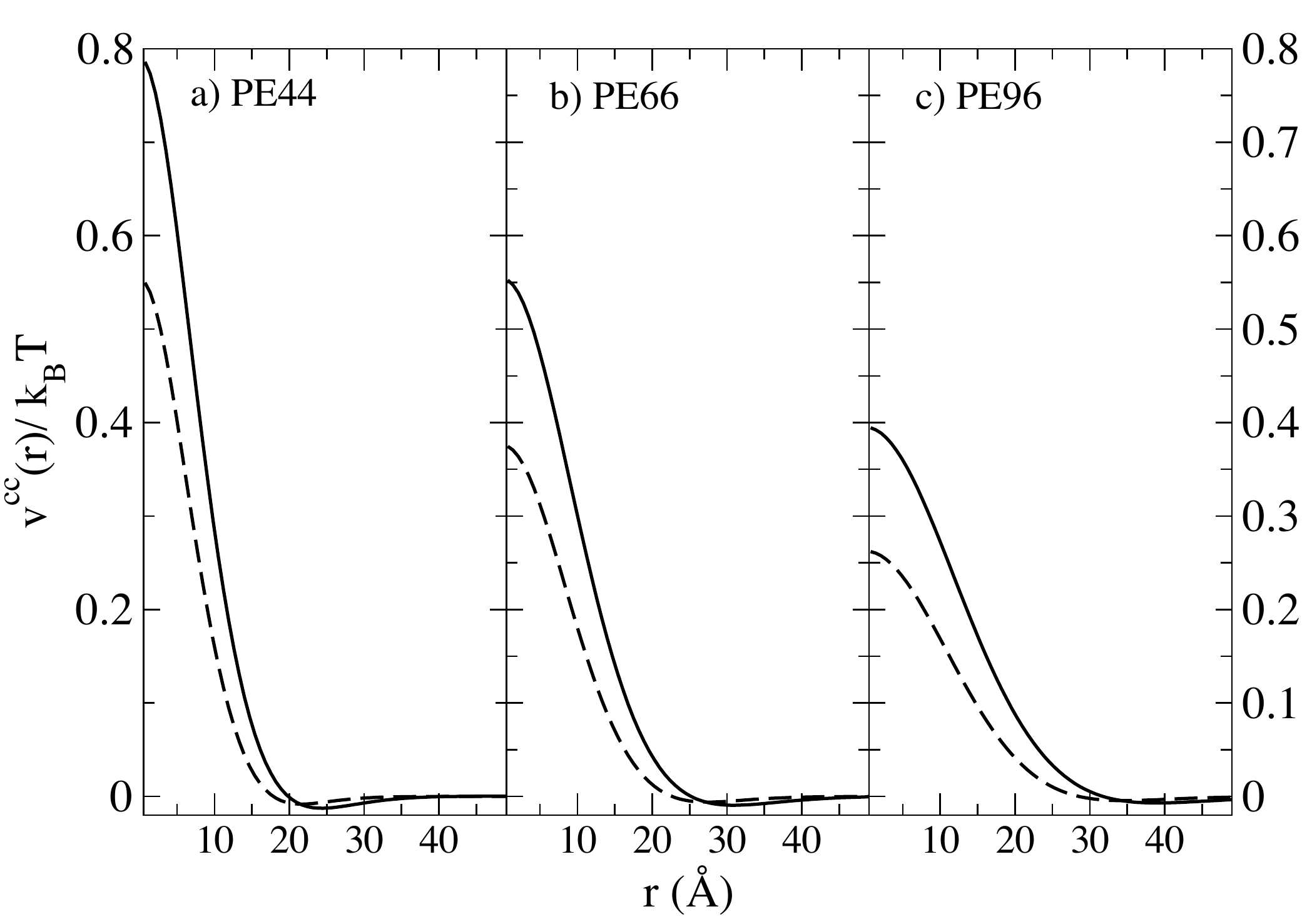}
\caption[Plot of analytical potential and potential of mean force]{The total pair potential, which is input into our MS-MD simulation, is shown for different chain lengths of polyethylene (solid line). For comparison, the dashed line represents the corresponding mean-force potential.}
\label{FG:potential}
\end{figure}

Finally, it is important to notice that the mean-force potential is not the correct input potential for the mesoscale simulations, as the two agree only for systems at low density,\cite{McQuarrie} however it is a good first guess in the iterative numerical procedure commonly adopted. In Figure \ref{FG:potential} we show for the systems investigated the direct potential, which is input to our simulation, and the mean-force potential. The two appear to be related, even if their numerical value and the extent of the attractive contribution is quite different.

\section{Mesoscale Simulation}
Once the formalism to describe analytically the structure of a liquid of polymers in a coarse-grained representation is solved, and the effective potential between coarse-grained units is derived, it is possible to run directly mesoscale simulations of the coarse-grained systems.
Our meso-scale simulations were performed by implementing classical MD simulations in the microcanonical $(N,V,E)$ ensemble. We simulated liquids of point particles interacting through the soft potential resulting from Eqs.(\ref{EQ:5.1})-(\ref{EQ:6.1}), which is repulsive at short range and slightly attractive at longer range. The interparticle potential differs depending on the chemical nature of the polymer investigated, as well as thermodynamic parameters of density and temperature. The systems we simulated included melts of
polyethylene (PE) with $N=44$ (PE44), $N=48$ (PE48), $N=66$ (PE66), and $N=96$ (PE96) monomers. Other systems were melts of  syndiotactic polypropylene (sPP),  head-to-head polypropylene (hhPP), isotactic polypropylene (iPP), and polyisobutylene (PIB), all with $96$ united atoms per chain.
Each simulation was performed on a cubic box with periodic boundary conditions. We used reduced units in our simulation, such that all the units of length were scaled by $R_g$ ($r^*=r/R_g$) and energies were scaled by $k_B T$. The density was fixed at the value reported in Table \ref{TB:SimParam}. Temperature and radius-of-gyration were utilized for dimensionalizing  the results obtained from the MS-MD simulations, after they were performed. 

\begin{table}[h!b!p!] 
\caption{Simulation Parameters for Polyolefin Melts}
\centering
\begin{tabular}{lcccc}
  \hline \hline
Polymer & Number of $CH_x$ moieties & T [K] & $\rho$ [sites/\AA$^3$] & $R_{g}$ [\AA]  \\
\hline 
PE44    &   44  &   400   &   0.0324    &   10.50 \\
PE48  & 48 &  448 & 0.0329 &  10.88 \\
PE66    &   66  &   448   &   0.0329    &   13.32 \\
PE96    &   96  &   453   &   0.0328    &   16.78 \\
sPP     &   96  &   453   &   0.0328    &   13.93 \\
hhPP    &   96  &   453   &   0.0336    &   13.54 \\
iPP     &   96  &   453   &   0.0328    &   11.34 \\
PIB     &   96  &   453   &   0.0357    &   9.68 \\
\hline \hline
\end{tabular}
\label{TB:SimParam}
\end{table}

The potential and its corresponding derivative entered as tabulated inputs to the program, and linear interpolation was used to determine function values not found in the supplied numerical data as the 
algorithm proceeds. In the initialization step, all particles were placed on a lattice, and each site was given an initial velocity pooled from a Mersenne Twister random number generator \cite{Frenkel}. 
Subsequently, the system was evolved using a velocity Verlet integrator. 

Equilibrium was induced in the ensemble by velocity rescaling. The desired temperature in the ensemble was attained by a quencing procedure, where velocities were forceably rescaled at regular intervals during the equilibration stage. We checked that the system had reached equilibrium when we observed a Maxwell-Boltzmann distribution of velocities, a steady temperature, a stabilized Boltzmann $H$-theorem function, and a decayed translational order parameter. At this stage, velocity rescaling was 
discontinued. We monitored the fluctuating steady temperature during the production stage to be sure that the system remains in equilibrium. During the production stage trajectories were collected 
over a traversal of $\sim8R_g$.

A typical mesoscale simulation consisted of $\sim3000$ particles, evolving for about $30,000$ computational steps during $\sim4$ hours, and performed on a single-CPU workstation. We stress that our mesoscopic simulations represent an underestimate in the computational time since these have not been subjected yet to a parallelization. 
The large-scale ($R_g$ lengthscale) pair distribution function resulting from the mesocale simulation has been shown to be in quantitative agreement with the one obtained from a full UA simulation, so that no precision is lost on the large scale during the coarse-graining procedure.\cite{YAPRL,YABLEND,SAM2006}
The mesoscale simulation, however, required a much smaller computational power than the full UA simulation. UA-MD simulations of polymer melts against which our MS-MD simulations are compared used $\approx 1,600$ chains with $N=96$ units, where $ 1 \ ns$ of simulation required either $9 \ h$ of computer time on $512$ processors of Sandia's ASCI Red Intel cluster or $25 \ h$ of computational time on $64$ processors on a Cplant DEC alpha cluster.\cite{MONDE,JARAM,HEINE} The reduced computational time of the MS-MD allowed for  improved precision in the large-scale regime as it was possible to increase the number of particles and the box size, with respect to the UA-MD, without dramatically affecting the computational time. 

In this paper we test our multiscale procedure by comparing the resulting monomer total distribution function against the  UA-MD simulation just described. We call this simulations the ``full UA-MD". Those simulations were performed by  Mondello et al.,\cite{MONDE} Jaramillo et al.,\cite{JARAM} and Heine et al.\cite{HEINE}.

\begin{table}[h!b!p!] 
\centering
\caption{Comparison of Mesoscopic Simulations (MS-MD) with United Atom Simulations (UA-MD). Number of molecules $n_{MS}$ in a box of length $L_{MS}$ in MS-MD; number of molecules $n_{UA}$ in a box of length $L_{UA}$ in the full UA-MD}
\begin{tabular}{ccccc}
  \hline \hline
System & $n_{MS}$ &  $L_{MS}$[\AA] & $n_{UA}$ & $L_{UA}$ [\AA] \\
\hline
PE44    & 4000    &   175.7978  &   100     &   51.4033 \\
PE48 & 4096 & 181.6127 & 400 & 83.634\\
PE66    & 5324    &   220.2708  &   100     &   58.5530 \\
PE96    & 6859    &   271.7286  &   100    &   167.270 \\
sPP     & 4394    &   234.2482  &   1600    &   167.27 \\
hhPP    & 4096    &   227.5623  &   1600    &   165.966 \\
iPP     & 3375    &   214.5239  &   1600    &   167.27\\
PIB     & 3375    &   208.6244  &   1600    &   162.676 \\
\hline 
\end{tabular}\\
\label{TB:MSUA}
\end{table}

To capture properties on the monomer scale, the multiscale procedure requires information on the local scale, which must be extracted from simulations with atomic level resolution, here UA-MD. However, the UA simulation input to a multiscale modeling procedure only requires a number of polymers of the order of $\sqrt{N}$, which is much smaller than the number of polymers simulated in the full UA-MD, as we will discuss further in the paper. Henceforth we refer to the UA-MD simulation input to the multiscale procedure as the ``local UA-MD".

\section{Building the full structural description at the monomer scale using a multiscale approach}

In a multiscale procedure, simulations are performed on a hierarchy of models where the same system is coarse-grained at different levels of resolution. The success of the method depends on the extent to which the system properties are modulated around distinct length scales. 
In the description of a polymer melt, the two lengthscales that are involved are the monomer segment length and the polymer radius-of-gyration. As it is shown in the left panels of Figures \ref{FG:5}-\ref{FG:7} for the polymers considered in this study the peak on the local scale and the global properties are well separated. The multiscale procedure would not work if the local and global features superimpose, which could happen for short chains of stiff polymers.
For those systems, however, the full UA-MD simulations are not too time consuming and there is no need to apply coarse-graining or to adopt a multiscale procedure. 

In our mesoscale simulations polymers are represented as point particles interacting through a soft potential of the range of the polymer size. i.e. $R_g$. The MS-MD provide information for any lengthscale of the order and larger than the polymer radius-of-gyration. It should be noted that neither mesoscale nor UA simulations can accurately describe the system at a radius larger than half their simulation box length. However, because the box size can be considerably larger for mesoscale simulations than for UA-MD, the MS-MD can describe large-scale properties more efficiently than UA simulations (see Table \ref{TB:MSUA}).  

On the local scale, however, mesoscale simulations do not provide any information. To recover the structure of the liquid at lengthscales smaller than or equal to $R_g$, it is necessary to combine MS-MD with UA-MD simulations. These UA-MD, however, only need to provide information on the local scale, and they are run on a small number of molecules, of the order of $\sqrt{N}$,   and are performed for a contained number of simulation steps. Even if UA-MD simulations have to be performed to provide the local information, the multiscale procedure efforts a net computational gain with respect to directly running a ``full" UA-MD.

\begin{figure}[]
\centering
\includegraphics[scale=.8]{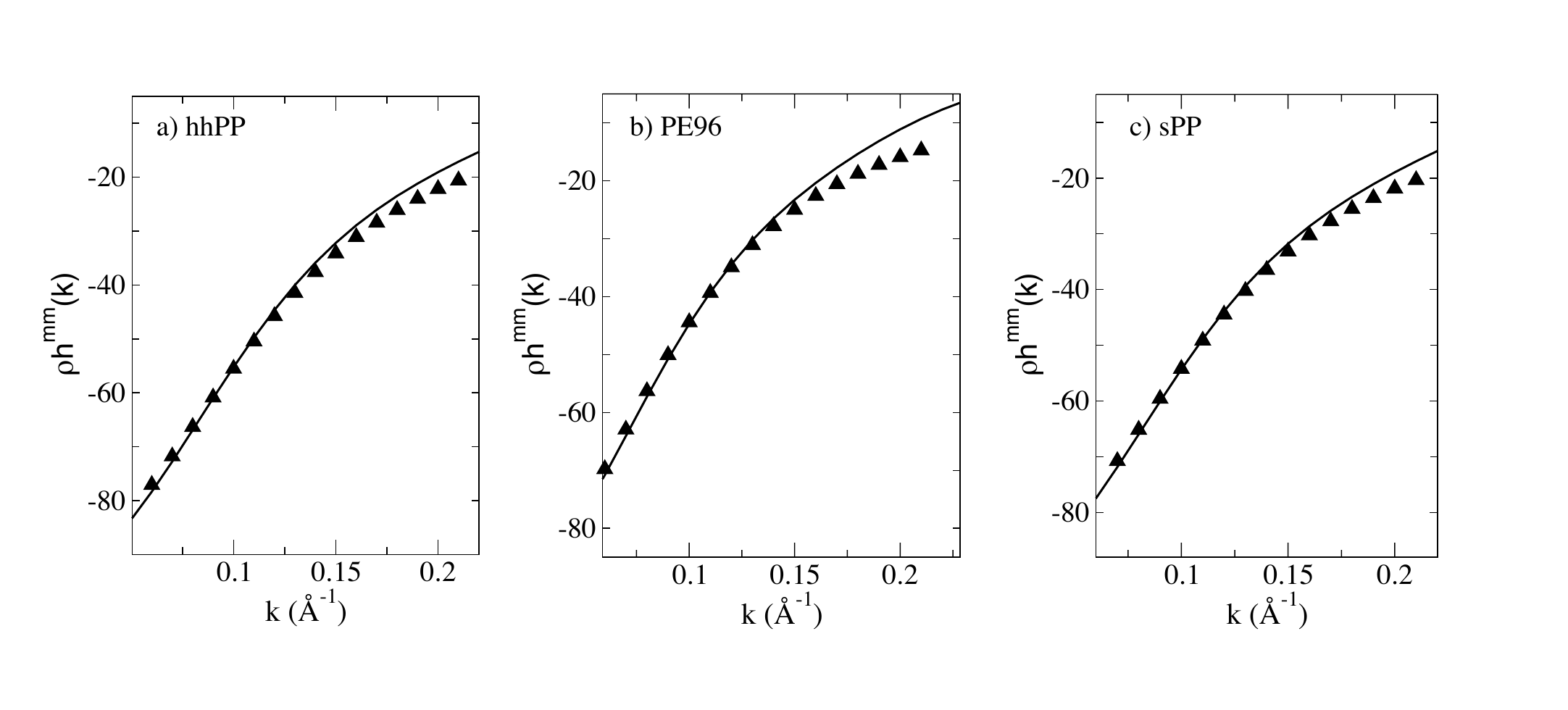}
\caption[Plot of $h^{mm}(k)$ theory simulation]{ Plot of $h^{mm}(k)$ for different polymer melts. Solid line depicts $h^{mm}(k)$ calculated using  Eq.(\ref{EQ:3}). Symbols represent data points from full UA MD simulation.}
\label{FG:1}
\end{figure}

\begin{figure}[]
\centering
\includegraphics[scale=0.6]{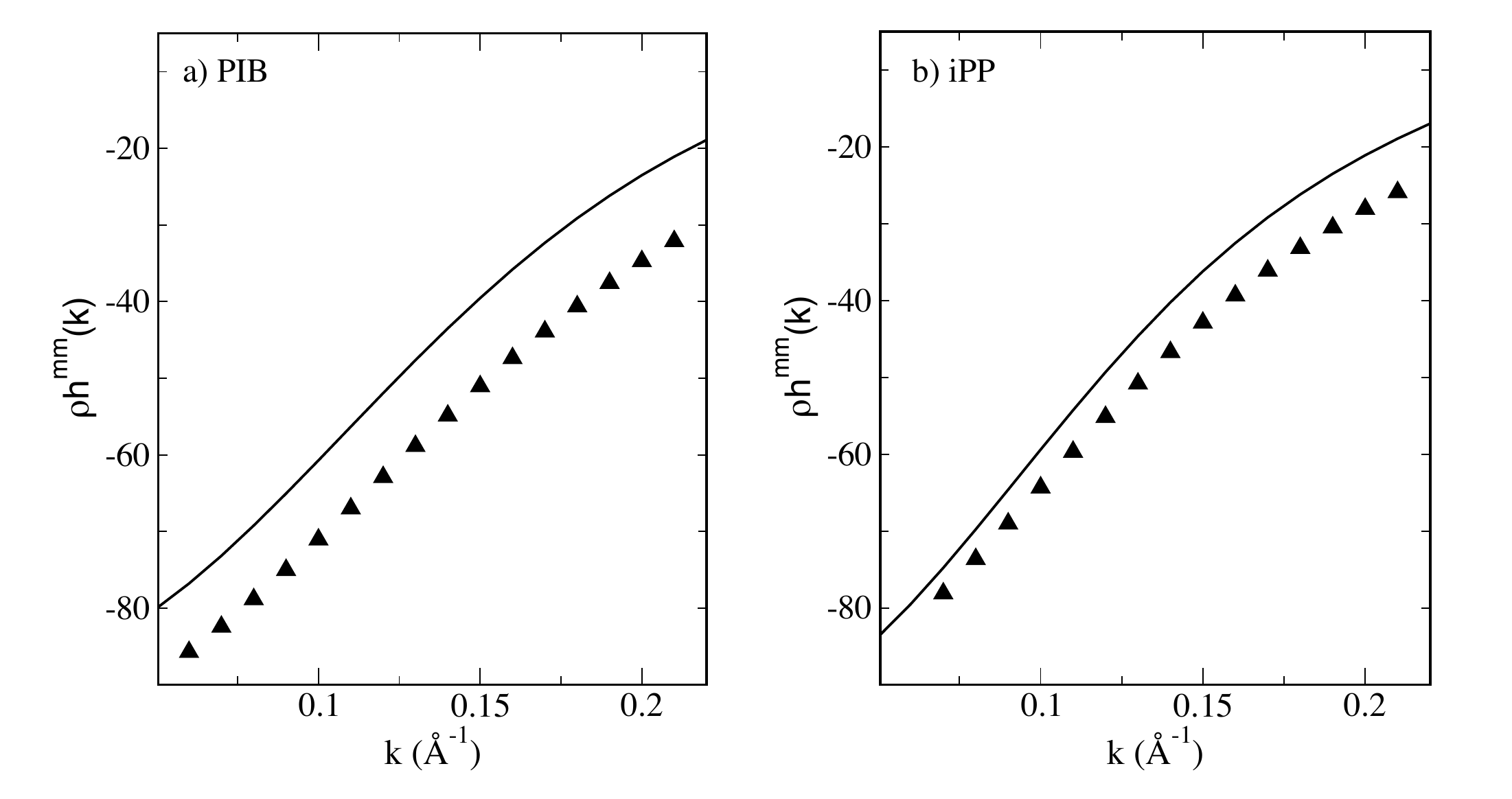}
\caption[Plot of $h^{mm}(k)$ theory simulation]{Plot of $h^{mm}(k)$ for different polymer melts. Solid line depicts $h^{mm}(k)$ calculated using  Eq.(\ref{EQ:3}). Symbols represent data points from full UA MD simulation.
}
\label{FG:2}
\end{figure}

On the global scale, small $k$, the $\it{monomer}$ total distribution function is calculated  from the data of the mesoscale simulation, using the inverse of Eq.(\ref{EQ:2})
\begin{equation}
h^{mm}(k)=\left[\frac{\omega^{mm}(k)}{\omega^{cm}(k)}\right]^2h^{cc}(k) \ ,
\label{EQ:3}
\end{equation}
where mesoscale simulations provide $h^{cc}(k)$ while UA MD simulations are used to determine the intramolecular form factors,  $\omega^{mm}(k)$ and $\omega^{cm}(k)$.

Figure \ref{FG:1} and \ref{FG:2} show $h^{mm}(k)$ (solid line) calculated from the mesoscale simulation using 
Eq.(\ref{EQ:3}) for several of the polymer melts of Table \ref{TB:SimParam}, where each chain has $96$ $CH_x$ units, with $x=1,2,$ or $3$.
The calculated $h^{mm}(k)$ (solid line) is compared to the full UA MD simulation (symbols). As expected, the theory compares well to UA MD simulations for the small k range (large r), and begins to diverge as k increases. 
This is a consequence of the coarse-graining procedure, Eq.(\ref{EQ:3}). Since monomers can superimpose with the center-of-mass, $\omega^{cm}(k)$ approaches zero for large $k$, while $\omega^{mm}(k)$ has a finite value at contact due to the hard-core monomer-monomer repulsion. As a consequence the ratio: $[\omega^{mm}(k)/\omega^{cm}(k)]^2$ diverges at some large $k$ value. 

In general, $h^{mm}(k)$, calculated from Eq. (\ref{EQ:3}), using data from mesoscale simulations and the intramolecular structure factors from UA simulations, accurately captures the structural properties of the polymer for small values of $k$ (global structure), as it is shown in Figure \ref{FG:1}. 
However, the same quality of agreement observed between mesoscale simulations and data of the monomer total correlation function for most of the cases is not observed in two specific polymer structures, namely for iPP and for PIB, see Figure \ref{FG:2}. It is known that both PIB and iPP favor highly packed intramolecular structures because of the symmetry in the position of the sidechains in the monomeric unit\cite{SAM2006}. Moreover, because of the presence of two side-groups in a PIB monomer, the number of monomers in a PIB chain containing $96$ united atoms, is equal to $24$ monomers. With such a small number of monomers the PIB sample can hardly exhibit Gaussian statistics for the intramolecular distribution. Because the assumption of having a Gaussian statistics is the starting point in our approach,\cite{SAM2006,YABLEND,YAPRL} this accounts for the discrepancies between the coarse grained values and UA MD data that we observe for these systems.

\section{Combining United Atoms and Coarse-Grained simulations: the crossover lengthscale}

Since the mesoscale simulation only captures global properties, it is necessary to determine the length at which local, intramolecular effects become significant and cannot any longer be discarded. This defines the crossover lengthscale for the UA simulation at which data from UA-MD and MS-MD simulations are combined. 

The extent to which intramolecular effects remain important on large lengthscales depends on the length of the polymer and its flexibility.
If the length of the polymer is constant, stiffer polymers span a larger volume and therefore have a higher (lower) number of inter- (intra)molecular contacts than their more flexible counterparts. Furthermore, longer chains of polymers with identical chemical structure have a higher statistical number of interpenetrating molecules and a higher (lower) number of inter- (intra)molecular contacts than shorter chains. 
The iPP and PIB samples presented here are characterized by efficient intramolecular packing, which is due to the particular monomeric structure: for these chains intramolecular interactions are dominant over intermolecular ones for an extended range of lengthscales. 

\begin{figure}[]
\centering
\includegraphics[scale=0.6]{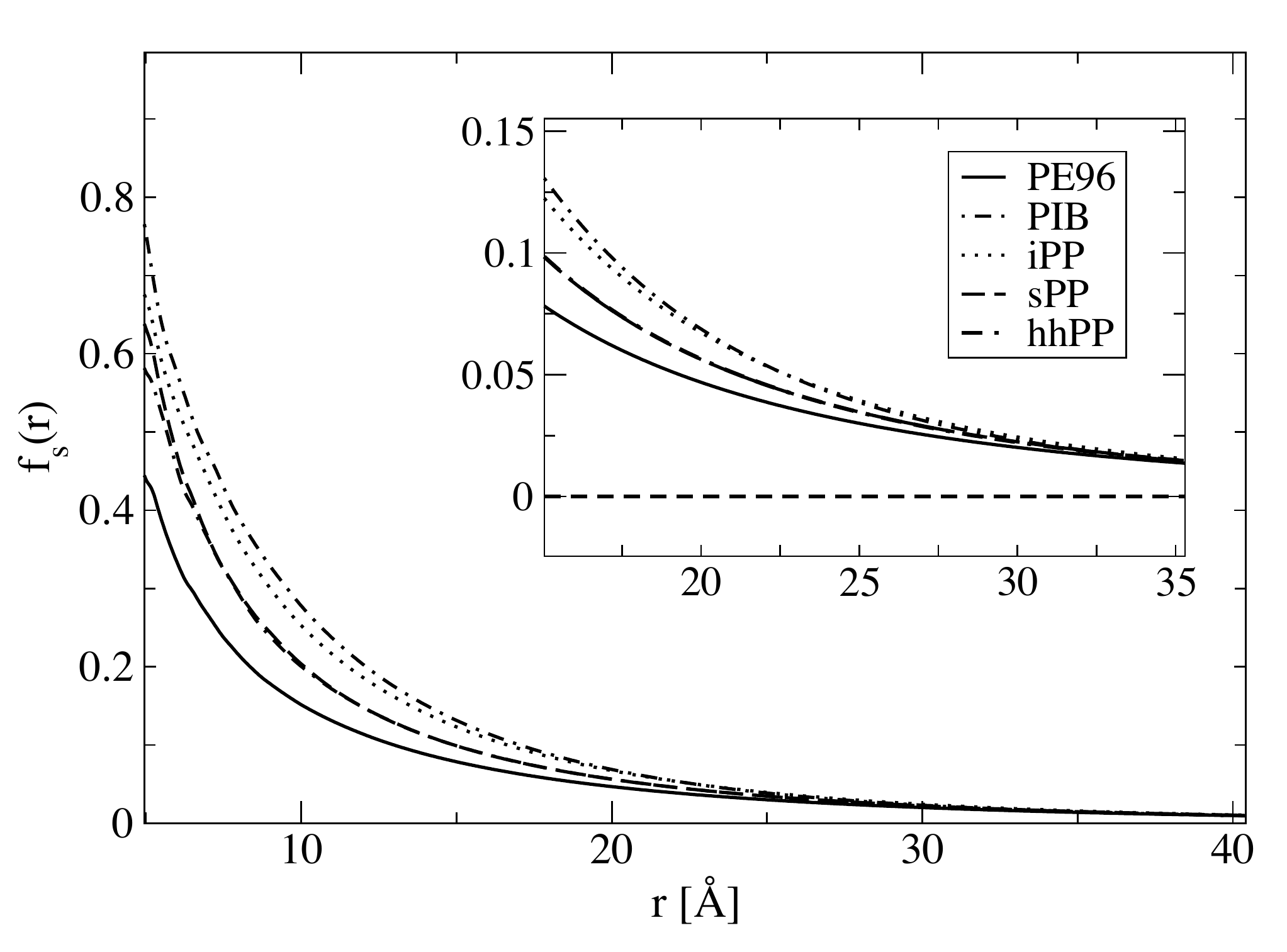}
\caption[plot of $f_s(r)$ for different polymer melts]{Fraction of intramolecular contacts, $f_s(r)$, for melts of polymers with different monomer architectures.}
\label{FG:3}
\end{figure}

The extent of intramolecular packing is quantified by calculating the ratio between the number of \emph{intra} and \emph{total} site/site contacts at a fixed radial distance, $r$,
\begin{equation}
f_s(r)=\frac{N_s(r)}{N_{total}(r)}
\label{EQ:8}\
\end{equation}

\noindent where $N_s(r)$ is the number of intramolecular contact sites defined as
\begin{equation}
N_s(r)=4 \pi \rho \int^r_0 r'^2\omega^{mm}(r')\,dr' \ .
\label{EQ:9}\
\end{equation}

\noindent The total number of sites in a given volume, $N_{total}(r)$, is given by
\begin{equation}
N_{total}(r)= \frac{4}{3} \pi \rho\,r^3 \, .
\label{EQ:10}\
\end{equation}

Figure \ref{FG:3} shows $f_s(r)$ vs. $r$ for different monomer architectures. As expected, iPP (dotted line) and PIB (dot-dashed line) show higher values of $f_s(r)$, whereas the stiffest molecule, PE (solid line), exhibits the lowest value at a given radius. 

Following the reasoning presented at the end of this section, we assume that a fraction of intramolecular contacts equal to $f_s(r)=0.025$, which corresponds to 2.5\% of the total sites involved in intramolecular contacts, defines the value of the crossover distance. This corresponds to a distance $r$ that is unique for a given polymer chain, and it defines the corresponding value in reciprocal space, $k=2\pi/r$, at which the two simulations have to be combined.  

\begin{figure}[]
\centering
\includegraphics[scale=0.8]{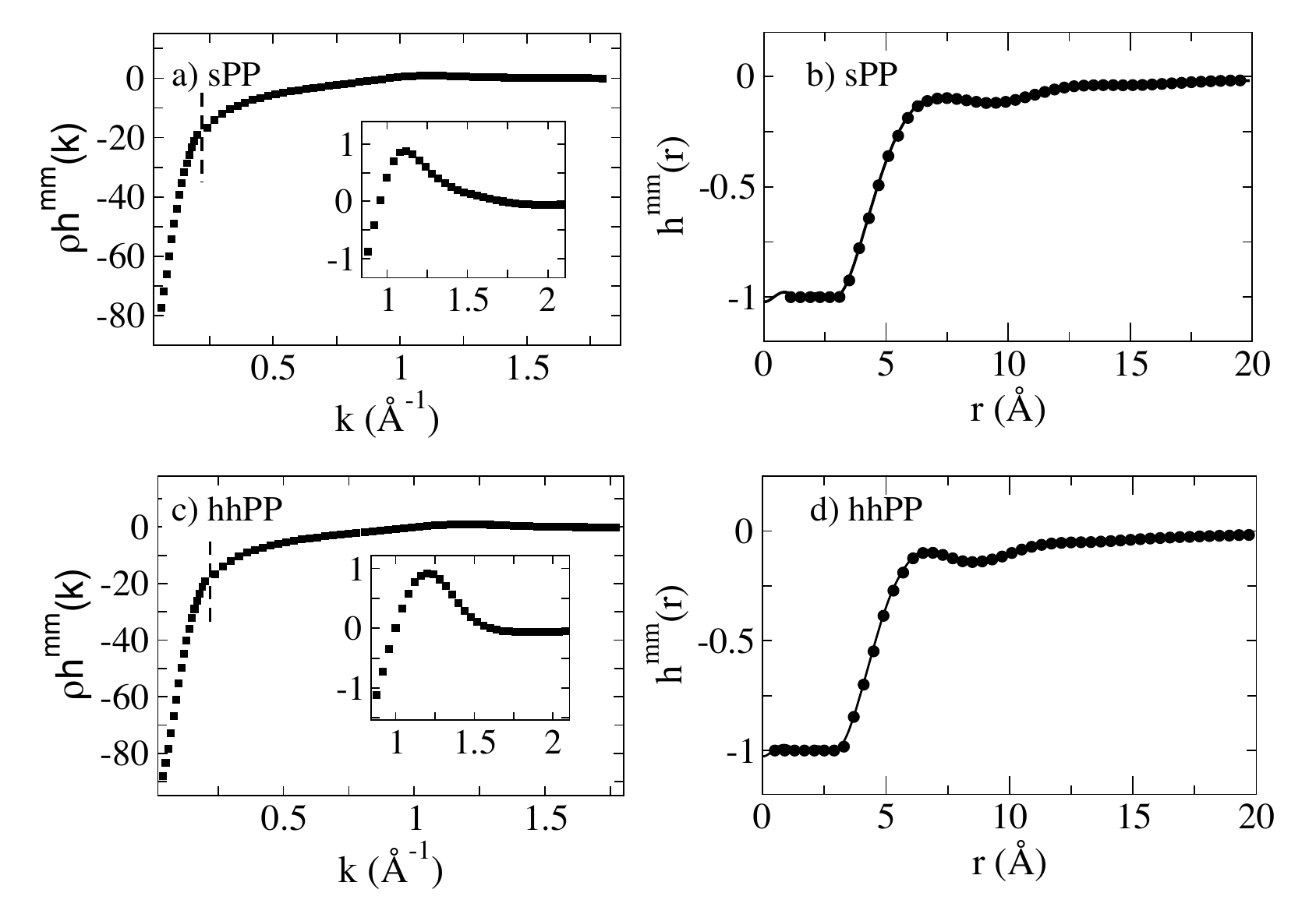}
\caption[Plot of $h^{mm}(k)$ total combining theory and UA simulation]{Left: Plot of the total correlation function, $h^{mm}(k)$, for polymer melts of (a) sPP and (c) hhPP, obtained by combining mesoscale and UA MD simulations. Mesoscale simulation depicts $h^{mm}(k)$ over the small k range whereas UA simulation provides data over the large k range. The dashed line indicates the point at which the two simulations were combined. The inset depicts the local peak. Right: Plot of the related $h^{mm}(r)$, the total correlation function in real space for (b) sPP and (d) hhPP. The solid line depicts $h^{mm}(r)$ calculated using our multiscaling approach and the open circles represent data points from UA MD simulations.
}
\label{FG:5}
\end{figure}

Figures  \ref{FG:5}-\ref{FG:7} display the comparison of the total correlation function obtained with our procedure (Eq.~(\ref{EQ:3})) and data from extended UA simulations.
The left panel of the Figures show the total correlation function in reciprocal space, $h^{mm}(k)$, for various polymer melts, (sPP, hhPP, iPP, and PIB) and for varying chain lengths of polyethylene (PE).
$h^{mm}(k)$ is obtained by combining the large scale data (small $k$) from the mesoscale simulations with the small scale data (large $k$) from united atom simulations. 
The vertical dashed line in Figures \ref{FG:5}-\ref{FG:7} indicate the radius at which the MS-MD is  combined with the UA MD simulation. Note that this occurs at an intermediate distance between the local peak and the global feature, so that neither information about the long range structure, nor the local structure is lost in extrapolating the connection between the two curves for $h^{mm}(k)$. 

\begin{figure}[]
\centering
\includegraphics[scale=0.8]{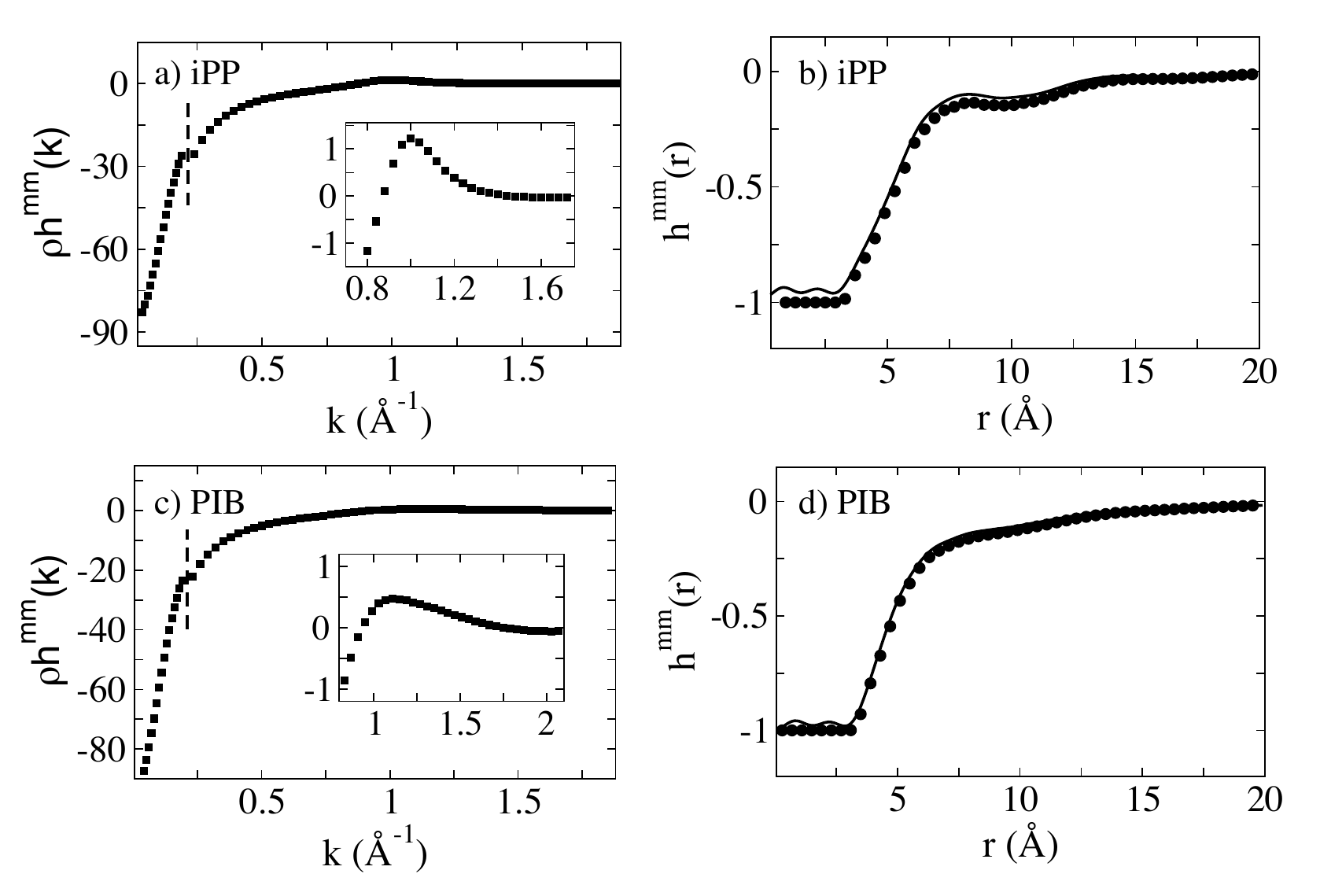}
\caption[Plot of $h^{mm}(k)$ total combining theory and UA simulation]{Same as in Figure \ref{FG:5} for  polymer melts of (a) iPP and (c) PIB}
\label{FG:6}
\end{figure}

The right panels of Figures \ref{FG:5}-\ref{FG:7}  show the total correlation function, $h^{mm}(r)$, for polymer melts, obtained by taking the Fourier transform of $h^{mm}(k)$. The function, $h^{mm}(r)$, provides a complete description of the liquid structure and thermodynamic properties. In performing the discrete Fourier transform of the multiscale total correlation function, a sampling step of $\Delta k = 0.01${\AA}$^{-1}$ \, was used, which defines the resolution of the transformed function. Since we are combining two separate data sets, $\Delta k$ must be large enough so that the Fourier transform is not affected by the small discontinuity at the point of intersection. As long as the interval of the discontinuity is of the same order as that of the sampling step there is no effect on the Fourier transform. 

The calculated values for $h^{mm}(r)$ using our multiscaling approach (solid line) are presented along with data from the full UA MD simulations (symbols). The proposed method gives results in good agreement with the full UA simulation data and correctly captures all of the relevant structural features of the liquid, including solvation shells and the correlation hole observed in polymers. 

\begin{figure}[]
\centering
\includegraphics[scale=0.9]{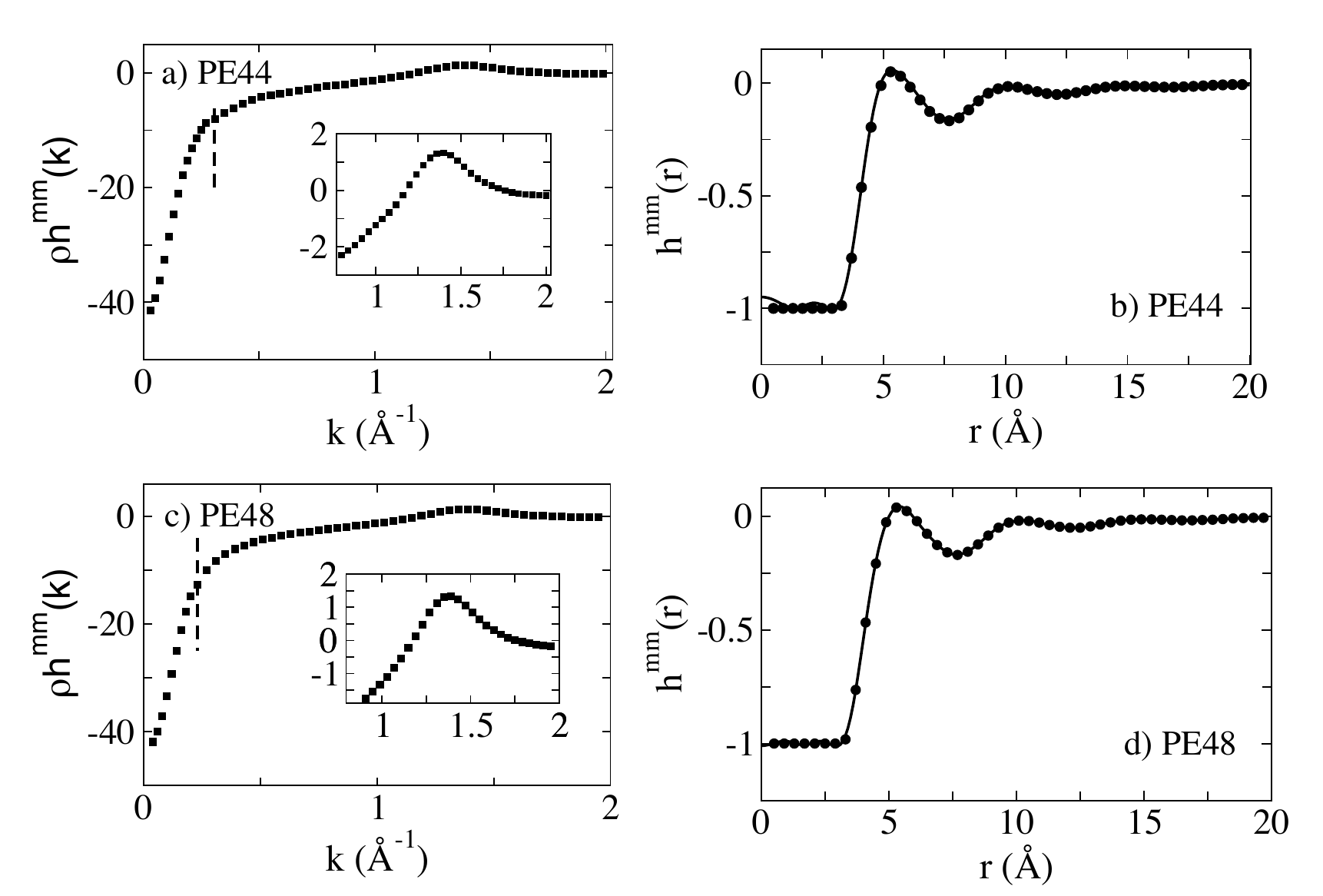}
\caption[Plot of $h^{mm}(k)$ and $h^{mm}(r)$ for polyethylene of different lengths]{Same as in Figure \ref{FG:5} for  polymer melts of (a) PE44 and (c) PE48}
\label{FG:7}
\end {figure}

\begin{figure}[]
\centering
\includegraphics[scale=0.9]{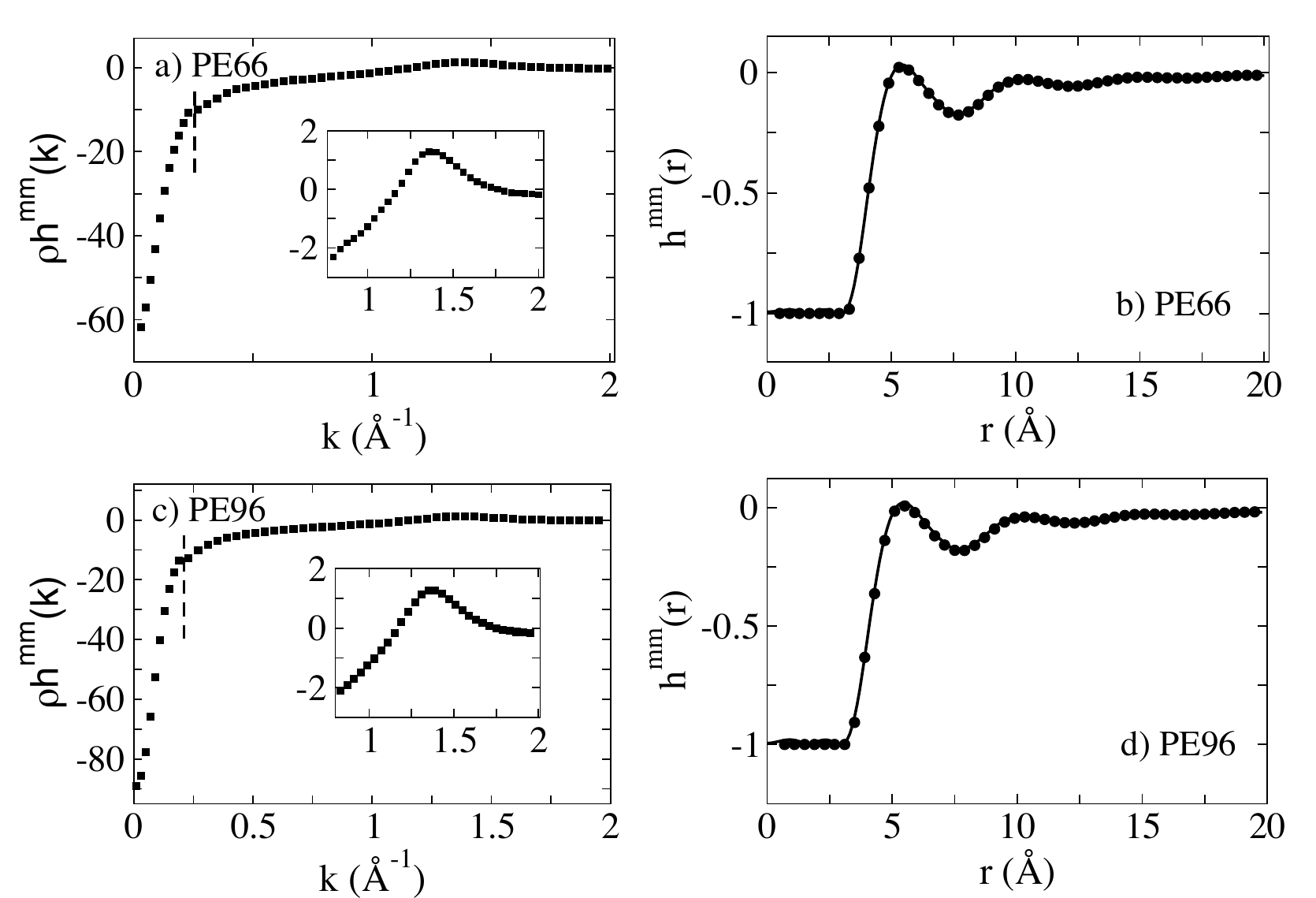}
\caption[Plot of $h^{mm}(k)$ and $h^{mm}(r)$ for polyethylene of different lengths]{Same as in Figure \ref{FG:5} for  polymer melts of (a) PE66 and (c) PE96}
\label{FG:7}
\end {figure}

Once this procedure is adopted, the iPP and PIB samples also give a good agreement with the full UA-MD simulation. Because  in these samples, and in PE44, the chains are short, the total correlation function still shows a small numerical error. The error is clearly displayed by the function in the $r < d$ regime, with $d$ the hard-core diameter of the repulsive inter-site interaction. This is the region of the excluded volume, and here it should be found that $h(r)=-1$. 

\begin{figure}[]
\centering
\includegraphics[scale=0.8]{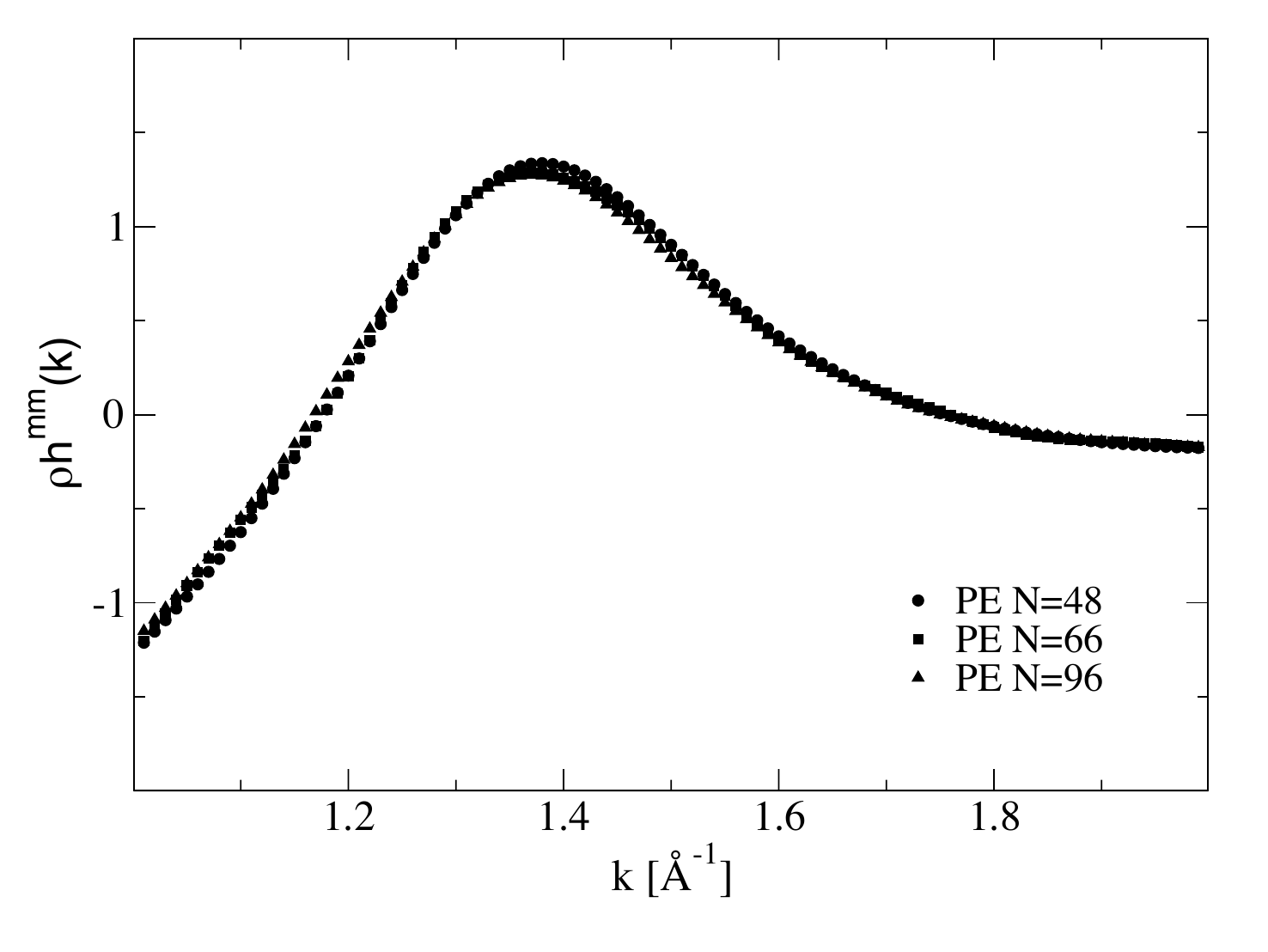}
\caption[Plot of the local peak from UA simulation]{Plot of the local peak for the total correlation function, $h^{mm}(k)$, for polymer melts of PE at increasing degree of polymerization, ($N=48, 66, 96$) from local UA simulations at T=448 K.}
\label{FG:8}
\end{figure}

The local chain structure, at large-$k$, is represented by the peak in the insets of Figures \ref{FG:5}-\ref{FG:7}, which has a shape that depends on the monomeric structure as well as on the thermodynamic parameters of density and temperature, as the local chain packing is affected by those quantities.  Figure \ref{FG:8} shows superimposed the local peaks for polyethylene chains with increasing degree of polymerization at constant temperature of 448 K. Because the chains have the same monomeric structure, and the thermodynamic parameters for each sample are close in value, the local peaks superimpose. This leads to a  further computational gain for the multiscale procedure, with respect to performing the full UA-MD simulation. Because the local properties are largely independent on the global scale properties for samples with long polymer chains, the multiscale procedure allows one to obtain the local scale properties for all samples just from one local UA-MD, which can be performed on a melt of short polymer chains, e.g. $N \approx 40$.

Once the pair distribution function is obtained from the multiscale procedure any thermodynamic quantity of interest can be numerically derived, including the equation of state, the virial coefficient, internal energy, compressibility, and others.\cite{McQuarrie}  Clearly the degree of precision of the multiscale total distribution function will affect the precision of the resulting thermodynamic quantities, so that care has to be taken in the choice of the crossover distance where local and global information is combined to optimize the agreement between UA-MD and multiscale data. 

The crossover distance, at which the UA-MD ceases to provide relevant information on the local scale, is determined from the chosen value of the fraction of intramolecular contacts. In all the calculations reported above we adopted the value $f_s(r)=0.025$. To determine this value,  we investigated the extent of agreement between the total correlation function calculated through the multiscale procedure and the one from the full UA-MD data as a function of $f_s$.

\begin{figure}[]
\centering
\includegraphics[scale=0.8]{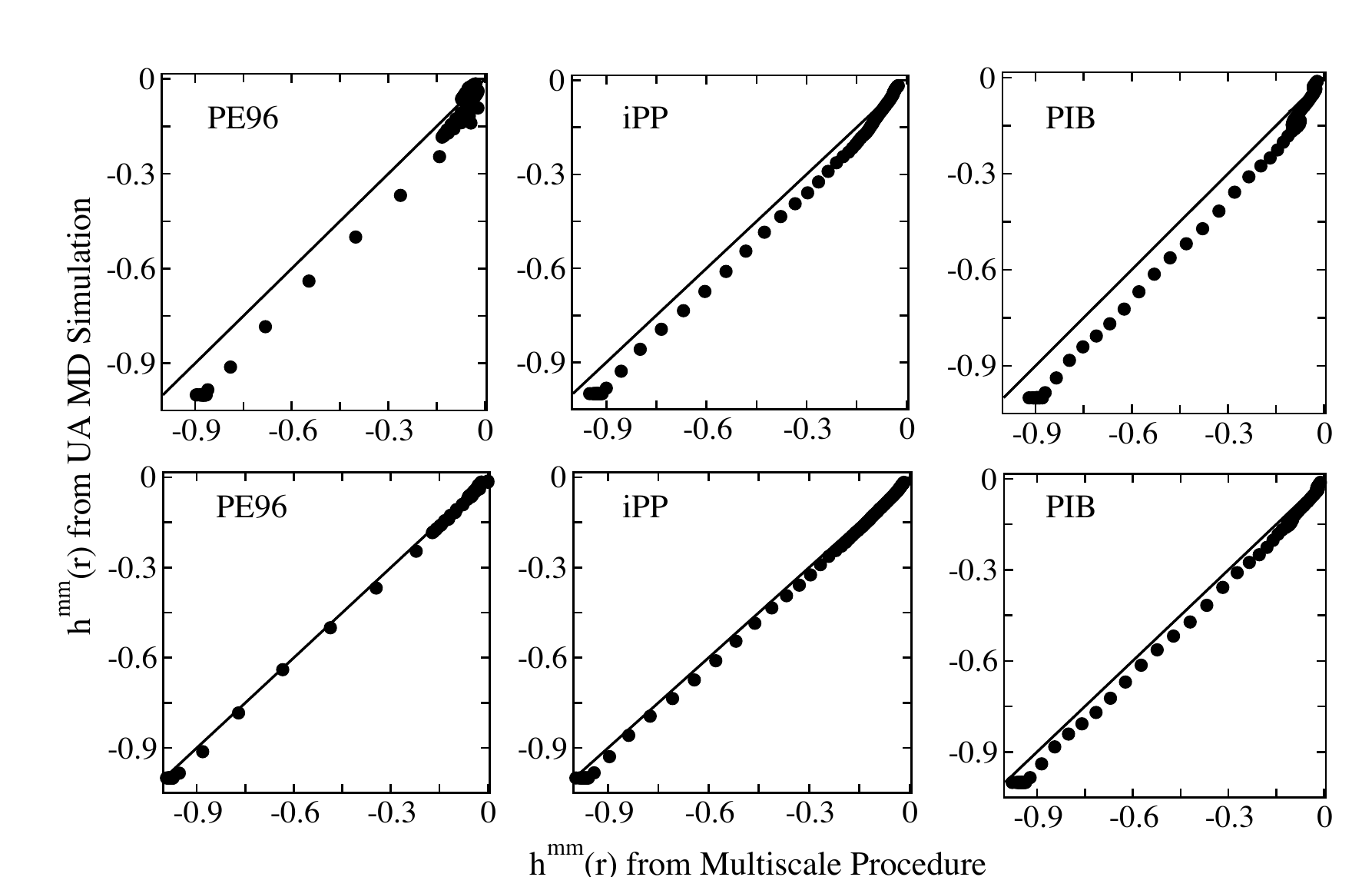}
\caption[correlation plot]{Correlation coefficient for $h^{mm}(r)$ obtained from the full UA-MD simulation and its theoretical value calculated with $f_s(r)=0.05$. Bottom panels: same as in top panels, but with $f_s(r)=0.025$.
}
\label{FG:4}
\end{figure}

Figure \ref{FG:4} shows a correlation plot of $h^{mm}(r)$ from UA simulation and $h^{mm}(r)$ calculated from our theory when local UA-MD data were combined with MS-MD data at a radius corresponding to $f_s(r)=0.05$ (top) and $f_s(r)=0.025$ (bottom). Shown are the correlation plots for the stiffest chain, PE, and for the two most flexible ones, PIB and iPP. The degree of correlation between theory and simulation increases as $f_s(r)$ decreases, as expected; however already at $f_s(r)=0.025$ there is a very high degree of positive correlation between the two. The correlation coefficients between theory and simulation for each polymer architecture are shown in Table \ref{TB:CorrCoeff}. As $f_s(r)$ decreases, the mesoscale picture of the liquid becomes increasingly accurate. However, $f_s(r)$ only becomes appreciably small for a large radius, $r$, and so there is a tradeoff between computational time in performing UA MD simulations and obtaining an accurate picture of the liquid structure.

Because the correlation coefficient is close to unity in all polymer systems tested for $f_s(r)=0.025$ and does not improve significantly for smaller values of $f_s(r)$, we propose that this value serves as a general rule of thumb for combining mesoscale and UA simulations. 

\begin{table}[h!b!p!]
\caption{Correlation coefficient for polyolefin melts}
\centering
\begin{tabular}{l||c}
    \hline
\em polymer  &  correlation coefficient \\ & at $f_s(r)=0.025$ \\
	\hline \hline
PE96      & 0.9998 \\
sPP       & 0.9999 \\
hhPP      & 0.9998 \\
iPP       & 0.9998 \\
PIB       & 0.9995 \\
\hline
\end{tabular}
\label{TB:CorrCoeff}
\end{table}

A possible different criterion of selection of the crossover lengthscale could be to defined the point in which the UA-MD curve intersect the MS-MD (see Figure \ref{FG:1}). This leads to values of the distance $r$ that are slightly larger than the ones established above, and for this reason requires more demanding local UA-MD, without adding precision to the calculated pair distribution functions. Furthermore, as mentioned above, for PIB and iPP, the curves never superimpose making an exact criterion of this sort for the combination of the two simulations unclear. By using the method adopted above of calculating the fraction of intramolecular contacts to determine the relevant crossover length scale, an exact distance, at which to combine the information, is readily obtained, and this distance does not depend on explicitly knowing the full UA MD simulation.

\begin{table}[h!b!p!] 
\centering
\caption{Total Number of Sites ($N_{sites}$) and Number of Molecules ($n$) included in a spherical volume of radius $r$, evaluated for $f_s(r)=0.025$ and $0.05$}
\begin{tabular}{ccccc}
  \hline \hline
System & $N_{sites}(0.025) $& $n(0.025)$ & $N_{sites}(0.05)$& $n(0.05)$ \\
\hline
PE96    & 2785    & 29    & 995    &  10 \\
sPP     & 3547    & 37    & 1412   &  15 \\
hhPP    & 3518    & 37    & 1511   &  16 \\
iPP     & 3931    & 41    & 1758   &  18\\
PIB     & 3729    & 39    & 1749   &  18\\
\hline \hline
\label{TB:SimCompare}
\end{tabular}
\end{table}

The value of $f_s(r)=0.025$ is reached for a distance $r$ that is polymer dependent. The number of polymer chains statistically comprised in the volume spanned by $r$ defines an average number of chains, $n$, that we need to consider in a local UA-MD simulation to produce good statistical information. This number can be calculated from Eq. (\ref{EQ:10}) using the crossover radius and dividing the number of particles so obtained by the number of $CH_x$, with $x=0,1,2,3$, units contained in a single chain. The number of sites, $N_{UA}$, and the number of polymer chains, $n$, contained in the volume defined by  the given radius of intramolecular interactions, calculated for value of $f_s(r)=0.025$, as well as for $f_s(r)=0.05$, are presented in Table \ref{TB:SimCompare}. 
The number of chains $n$ is also the statistical number of macromolecules that need to be considered in a local UA-MD simulation and is of the order of the number of chains interpenerating a ``tagged" polymer,  $\sqrt{N}$. This number of chains is one order of magnitude smaller than the number of molecules commonly used in the full UA simulations of polymer melts.

\section{Discussion and Conclusions}
In this paper we presented a new multiscale approach to simulate the structure of polymer melts in a computationally efficient way. We first identified the relevant lengthscales in which information has to be collected. For a liquid of uncharged homopolymer chains there are two lengthscales, which characterize all the structural properties of the macromolecular liquid, namely the monomer and the molecular radius of gyration. 

Molecular Dynamics simulations are performed for each system in the two different coarse-grained representations. At the monomer level, the liquid is  well represented through an United-Atom (UA) description, where each $CH_x$ unit, with $x=1,2,$ or $3$, corresponds to an effective unit. At the molecular level, the coarse-graining is obtained through a soft-colloid representation, where each polymer molecule is depicted as a point particle centered at the polymer center-of-mass, and interacting with other particles through a soft repulsive potential of the range of $R_g$.

The two coarse-grained descriptions are formally connected through an Ornstein-Zernike equation, solved under the condition that the coms are fictitious sites, while the united atoms are real sites. An analytical solution of the OZ equation in terms of the total correlation function between the two representations has been derived, and this equation is used to combine local and global scale information in the proposed multiscale modeling procedure.

Information on large scale properties is collected from the MS-MD, soft-colloidal, simulation, which extends to large-scale and covers long-time regimes, as the dynamics is speeded up because of the elimination of the internal degrees of freedom. The large-scale information has been shown to be quantitatively correct. However,  because  local degrees of freedom are averaged out, MS-MD simulations do not provide any information on local-scale properties. 

To obtain a description of the macromolecular liquid at the monomer lengthscale, MS-MD simulations have to be combined with local UA-MD simulations. A crossover lengthscale between the two representation has to be defined.
This is equivalent to establishing at which small lengthscale the coarse-grained description 
ceases to be valid. The local degrees of freedom are relevant
up to a characteristic lengthscale, which is different depending on the chemical structure of the monomer, but which can be evaluated using short-time small-box united atom simulations.  Once this crossover lengthscale is determined, it is adopted as the point at which global lengthscale information from MS simulations has to be combined with the local scale description from UA simulations, in reciprocal space.
The resulting monomer total distribution function is Fourier transformed from reciprocal to real space to
give the complete description of the system, through the total correlation function, $h(r)$.

We show that the resulting function, tested against data obtained from a full UA-MD simulation, exhibits quantitative agreement.
Furthermore, a comparison of the needed computational time and resources of the multiscale procedure, against the requirements of the full UA simulation for the same system, highlights several favorable points. 

In general, simulations of liquids necessitate the inclusion of a high number of molecules, $n$,  to achieve a good statistical description, minimize finite-size effects, and approximate the thermodynamic limit ($n \rightarrow \infty$  and $V \rightarrow \infty$  while density, $\rho=n/V$, is constant). Moreover, macromolecules usually comprise a large number of monomers  $N$, so that the overall computational time to simulate polymeric liquids increases rapidly, scaling as  $(nN)^2$, if we assume the conventional scaling of computational steps with the square of the number of particles. This feature poses some limitations on the total number of polymers, their length, or the longest timescale that can be simulated.

Other techniques can be adopted to decrease the exponent in the scaling of the computational time, but these procedures usually cannot be translated into parallel calculations. Even with the most powerful resources, long-time UA-MD simulations of polymer melts usually can treat a maximum of 500 chains, for polymers comprised by less than 100 monomers. However, the best statistical description is reached for simulations of thousands of molecules, which is very time consuming and impossible to perform for long polymers.

A multiscale procedure is advantageous, because the mesoscale simulation provides the information on the large spatial and long-time scale, so that microscopic UA simulations can be limited in the number of particles that are monitored and in the computational time during which data are recorded. 

In our procedure the number of molecules to be included in the UA simulation is of the order of the number of chains interpenetrating a single macromolecule (O$(\sqrt{N}$)), which is about one order of magnitude smaller than the number of particles needed if we perform a full UA simulation. For example, if we assume a scaling proportional to the square of the number of particles, the proposed procedure requires only $n^2=N$ soft colloidal particles to calculate $h^{cc}(k)$ and only $(nN)^2=N^3$ number of units in the local UA simulation, which is a dramatic gain with respect to the UA scaling of $(nN)^2$ where $n >> N$ for the full simulation.

If we assume a different scaling with the number of particles, for example because of adopting a more advanced sampling rule, the multiscale approach is still more convenient in the number of computational steps than the full UA MD simulations, since the gain in computational time when comparing the two procedures is due to the change in the total number of molecules to be tracked in the UA simulation, i.e. from $n \gg N$ in the full UA simulation to $n \propto \sqrt{N}$ in the local UA simulation.

Finally, computational time is further reduced when a family of homologous chains that only differ because of the degree of polymerization are under study: in this case local UA-MD only have to be performed for the shortest chain, as discussed in the previous section.

Having reduced the computational time necessary to obtain local information, the global scale description results from a mesoscale simulation, which can include  a large number of particles and large box size without dramatically increasing the computational time, thus improving the statistics of the system on the global scale. Table \ref{TB:MSUA} shows the comparison between the total ensemble number ($N$) and simulation box length ($L$) for UA simulations and MS simulations of different polyolefines.  

For example, the full UA simulation for polyethylene with 66 united atoms includes 100 polymer chains, while our MS simulation includes over 5,000 polymers as soft colloids, providing a mesoscale liquid structure which is identical in the two simulations and a large scale description with improved statistics. This reinforces the point that only small UA simulations need to be performed to capture local properties, whereas the global properties may be deduced using a mescoscale approach which greatly extends the capabilities for computer simulations of macromolecules. The possibility of simulating systems on very large lengthscales while including a large number of particles can be useful in treating systems with diverging correlation lengths, such as mixtures approaching their spinodal decomposition.

In conclusion, Figures \ref{FG:5}-\ref{FG:7} show that our method of combining mesoscale simulations at the coarse grained level with small UA simulations accurately captures both the large and small scale structure of a polymeric melt while remaining computationally advantageous. The approach is analytical and uses no adjustable parameters, and is therefore applicable to a large number of different polymer systems of varying chain length, density, and with different local chemical structure. As an example in this paper we investigate polyethylene chains with increasing degree of polymerization, as well as polymer chains with different monomeric structure. The same multiscale procedure is used for all of them, providing efficient calculations of monomer static properties across the many lengthscales of interest.

 \section{Acknowledgements}
We acknowledge the support of the National Science Foundation through grants DMR-0509808 and DMR-0804145.

\newpage
\end{document}